\documentclass[12pt]{iopart}

\expandafter\let\csname equation*\endcsname\relax
\expandafter\let\csname endequation*\endcsname\relax
\usepackage{amsmath,amssymb,amsthm}
\usepackage{txfonts,newtxmath}
\usepackage{braket}
\usepackage{easybmat}
\usepackage[colorlinks=true,citecolor=blue,urlcolor=blue]{hyperref}
\usepackage{times}
\usepackage{braket}
\usepackage{color}
 \usepackage{cite}

\newcommand{\be}{\begin{equation}}
\newcommand{\ee}{\end{equation}}
\newcommand{\ba}{\begin{eqnarray}}
\newcommand{\ea}{\end{eqnarray}}

\usepackage{hyperref}
\usepackage{orcidlink}
\usepackage{url}

\newcommand{\proj}[1]{\ket{#1}\bra{#1}}

\usepackage{graphicx}
\usepackage{bm,bbm}%
\usepackage{braket}
\usepackage{enumerate}
\usepackage{booktabs,multirow}
\usepackage{mathtools,bbding}
\usepackage{microtype}
\usepackage{mathrsfs}

\usepackage[most,breakable]{tcolorbox}

\newcommand{\nc}{\newcommand}
\newcommand{\rnc}{\renewcommand}

\rnc{\thesection}{\arabic{section}}
\rnc{\thesubsection}{\thesection.\arabic{subsection}}
\rnc{\thesubsubsection}{\thesubsection.\arabic{subsubsection}}

\usepackage{etoolbox}
\hypersetup{linktoc=all}

\nc{\tcr}[1]{{\color{red} #1}}
\nc{\tcb}[1]{{\color{blue} #1}}

\nc{\psic}{\psi^{c}}
\nc{\two}[1]{\underline{2^{d-#1}}}
\rnc{\H}{\mathcal{H}}
\nc{\Hanc}{\mathcal{H}_{\text{anc}}}
\nc{\psianc}{\psi_{\text{anc}}}
\nc{\lampow}{\lambda^{1/d}}

\nc{\RR}{{{\mathbb R}}}
\nc{\CC}{{{\mathbb C}}}
\nc{\FF}{{{\mathbb F}}}
\nc{\NN}{{{\mathbb N}}}
\nc{\ZZ}{{{\mathbb Z}}}

\nc{\MIO}{{\text{\rm MIO}}}
\nc{\DIO}{{\text{\rm DIO}}}
\nc{\SIO}{{\text{\rm SIO}}}
\nc{\IO}{{\text{\rm IO}}}

\nc{\SEP}{{\text{SEP}}}
\nc{\NS}{{\text{NS}}}
\nc{\LOCC}{{\text{LOCC}}}
\nc{\PPT}{{\text{PPT}}}
\nc{\EXT}{{\text{EXT}}}
\nc{\OLOCC}{{\text{1-LOCC}}}
\nc{\SEPP}{{\text{SEPP}}}

\nc{\MC}{{\text{\rm MC}}}

\nc{\cE}{\mathscr{E}}

\rnc{\*}{\textup{*}}
\newcommand{\id}{\mathbbm{1}}
\newcommand{\openone}{\mathbbm{1}}
\renewcommand{\S}{\mathcal{S}}

\renewcommand{\P}{\mathcal{P}}

\nc{\MM}{\widetilde{\M}}
\nc{\Ml}{\M^{\leq}}

\nc{\mleq}{\preceq}
\nc{\mgeq}{\succeq}

\nc{\ox}{\otimes}

\nc{\wt}{\widetilde}

\nc{\SDP}{\text{\rm SDP}}

\nc{\cc}{{\circ\circ}}
\nc{\mnorm}[1]{\norm{#1}{[m]}}

\nc{\F}{\mathcal{F}}
\nc{\M}{\mathcal{M}}

\let\oldproofname\proofname
\rnc{\proofname}{\rm\bf{\oldproofname}}
\rnc{\qedsymbol}{{\color{gray!50!black}\rule{0.6em}{0.6em}}}

\newcommand{\bb}{\begin{equation}}
\newcommand{\bbb}{\begin{equation*}}
\newcommand{\eee}{\end{equation*}}
\begin{document}

\title[Quantum Advantage in Assisted Work Distillation]{Quantum Advantage in Assisted Work Distillation from Complementary Correlations}

\author{Chellasamy Jebarathinam$^{1,2}$\orcidlink{0000-0003-2632-2174}, Debarshi Das$^3$ and Ray-Kuang Lee$^{4,5,6,*}$\orcidlink{0000-0002-7171-7274}}

\address{$^1$ Department of Physics and Center for Theoretical Physics, National Taiwan University, Taipei 106319, Taiwan}
\address{$^2$ Center for Quantum Science and Engineering, National Taiwan University, Taipei 106319, Taiwan}
\address{$^3$ Department of Physics, Shiv Nadar Institution of Eminence, Gautam Buddha Nagar, Uttar Pradesh 201314, India}
\address{$^4$ Institute of Photonics Technologies, National Tsing Hua University, Hsinchu 30013, Taiwan}
\address{$^5$ Center for Theory and Computation, National Tsing Hua University, Hsinchu 30013, Taiwan}
\address{$^6$ Center for Quantum Science and Technology, Hsinchu 30013, Taiwan}
\address{$^*$ Author to whom any correspondence should be addressed.}
\ead{rklee@ee.nthu.edu.tw}

\begin{abstract}
We establish an operational bridge between microscale thermodynamic work extraction and the foundational characterization of nonclassical correlations in bipartite quantum systems. Within the resource theory of thermal operations, we analyze a generalized framework of assisted work distillation where the trusted receiver's system is divided into separate, independent particle storage pools. We demonstrate that concurrent information-theoretic correlations measured across sets of mutually unbiased bases (MUBs) act as an explicit, quantifiable energy resource for local work extraction. Based on this paradigm, we derive a novel class of work-based Einstein-Podolsky-Rosen (EPR) steering inequalities whose violations certify a one-sided device-independent quantum thermodynamic enhancement. Furthermore, by restricting the dimension of the untrusted party under a one-sided semi-device-independent (1SSDI) trust framework, we introduce dynamically scaling work inequalities capable of witnessing superunsteerability. This allows us to activate and certify the hidden thermodynamic utility of weakly correlated entangled states that are otherwise completely invisible in standard, unconstrained scenarios. 
\end{abstract}


%\submitto{\QST}
\maketitle

\section{Introduction}

The intersection of quantum information theory and quantum thermodynamics demonstrates that nonclassical features can overcome classical restrictions in microscale thermodynamic tasks~\cite{janzing2000thermodynamic,brandao2013resource,Goold2016TopicalReview,Vinjanampathy2016review,Myers2022quantum,Cangemi2024quantum,HsiehGessner2026}. Within the resource theory of thermal operations (TO), quantum features can be actively harnessed to drive microscopic quantum devices~\cite{Brandao2015PNAS, Lostaglio2015Coherence, NgWoodsReview2018,Myers2022quantum}. For instance, EPR steering~\cite{EPR35,Schrodinger1935steering,WJD07} can empower a quantum Maxwell's demon to optimize work extraction from microscopic heat engines~\cite{BLS19,Ji2022spin,Cangemi2024quantum}. 

In such an asymmetric setting, a protocol known as \textit{assisted work distillation} has already been formalized by Morris \etal~\cite{MLA19}, where two observers share a bipartite state $\rho_{AB}$. While Bob is restricted to local TO at a background inverse temperature $\beta$, Alice possesses unconstrained operational capabilities and can send one-way classical communication to assist him. Under this paradigm, free TO on a system $S$ with Hamiltonian $H_S$ are completely-positive trace-preserving (CPTP) maps $\mathcal{E}$ of the form:
\begin{align}\label{TO}
	\mathcal{E}(\rho)=\Tr_{E}\left(U_{SE}\left(\rho_S\otimes\gamma_E\right)U_{SE}^\dagger\right),
\end{align}
where $U_{SE}$ is a global energy-conserving unitary satisfying $\left[U_{SE},H_S\otimes\id_E+\id_S\otimes H_E\right]=0$, and $\gamma_E={Z}_E^{-1}e^{-\beta H_E}$ represents the environmental Gibbs state.

The distillable work $W(\rho_B)$ from Bob's subsystem equates asymptotically to the change in non-equilibrium free energy~\cite{brandao2013resource,horodecki2013fundamental,Skrzypczyk2014work}:
\begin{equation} \label{workisfreeenergy}
    W(\rho_B) \equiv \Delta F(\rho_B) = \frac{1}{\beta} S(\rho_B\|\gamma_B) ,
\end{equation}
where $S(\rho\|\gamma)=\Tr\left(\rho \log \rho - \rho \log \gamma\right)$ is the relative entropy of athermality. 
When Alice applies a local positive operator-valued measurement (POVM) $\{\Pi_{A,i}\}$ with outcome probabilities $p_i=\Tr\left[\rho_A \Pi_{A,i}\right]$, Bob gains access to the conditionally prepared ensemble $\{p_i, \tilde{\rho}_{B,i}\}$, where:
\begin{align} \label{rho tilde}
	\tilde{\rho}_{B,i}=\frac{1}{p_i}\Tr_{A}\left[\left(\Pi_{A,i}\otimes\id_B\right) \rho_{AB}\right].
\end{align}
To optimize this cooperative distillation process, the work of assistance is defined as~\cite{MLA19}:
\begin{align}\label{workass1}
	W_{a}^{B|A}(\rho_{AB}):=\max_{\{\Pi_{A,i}\}}\frac{1}{\beta}\sum_i p_i S(\tilde{\rho}_{B, i}||\gamma_B),
\end{align}
which can be split into independent and classically correlated components~\cite{MPZ18,MLA19}:
\begin{align}\label{workass4}
W_a^{B|A}(\rho_{AB})=\frac{1}{\beta}\left(S(\rho_B||\gamma_B) + J^{\rightarrow}(\rho_{AB})\right),
\end{align}
where $J^{\rightarrow}(\rho_{AB})\coloneqq \max_{\{\Pi_{A,i}\}}\left(S(\rho_B)-  \sum_ip_iS(\tilde{\rho}_{B,i})\right)$ represents the Henderson--Vedral measure of classical correlations~\cite{HV01}. 
The clear operational separation of these components motivates the multi-setting setup investigated in this work, where the concurrent tracking of diverse measurement bases allows us to systematically expand the domain of certified thermodynamic advantages.

Motivated by thermodynamic demons operating under two complementary bases~\cite{BLS19}, the primary objective of this work is to generalize the framework of assisted work distillation to multi-setting measurement configurations. In a standard protocol, a localized quantum Maxwell's demon can typically extract work using only a single fixed measurement configuration~\cite{MLA19}. However, such a quantum Maxwell's demon does not utilize nonclassical correlations of shared bipartite state to demonstrate any quantum advantage. To overcome this limitation, we consider an operational scenario where Bob's storage system is divided into separate particle pools, each optimized to extract work from a distinct conditionally steered configuration. Within the asymmetric framework of EPR steering, we characterize the operational verification of nonclassicality for our thermodynamic demons through the optimization of the steerable weight of the prepared quantum assemblage~\cite{Skrzypczyk2014steerable,Cavalcanti2017review}.  
This optimization framework allows us to generalize assisted work distillation to multi-setting measurement configurations, establishing a novel class of work-based steering inequalities whose violations certify a one-sided device-independent quantum enhancement  via thermodynamic work extraction. Furthermore, by restricting the dimension of the untrusted observer, we introduce dimension-bounded steering inequalities capable of witnessing superunsteerability within the domain of unsteerable states~\cite{DBD+18}.

These bounds define strict thermodynamic constraints on local hidden state (LHS) models, complementing the foundational characterization of the entanglement hierarchy of LHS ensembles introduced by Fadel and Gessner~\cite{fadel2022entanglement}.
While our dimension-bounded steering inequalities demonstrate that entanglement—rather than EPR steering—serves as a more general resource for certified quantum advantage in assisted work distillation, complementary thermodynamic protocols have also been developed to witness entanglement and coherence directly via bounded heat exchange in memory-assisted settings \cite{deOliveiraJunior2025Heat}.

\section{Simultaneous correlations in mutually unbiased bases (SCMUB)}
\label{sec:scmub}

To ground this micro-thermodynamic framework within quantum information theory, we characterize the role of the optimized complementary measurement strategies using information-theoretic measures of nonclassical correlations~\cite{WMC+14,OZ01}.
If Alice projects her subsystem onto an arbitrary basis $\{\left|a_{i}\right\rangle _{A}\}$, the maximum information Bob can access regarding her localized outcomes is upper-bounded by the Holevo quantity:
\begin{equation}
\chi\big\{\rho_{AB}\big|\{\left|a_{i}\right\rangle _{A}\}\big\} \equiv S\Big(\sum_{i}p_{i}\tilde{\rho}_{B,i}\Big) - \sum_{i}p_{i}S(\tilde{\rho}_{B,i}), \label{eq:holevo_unified}
\end{equation}
which effectively denotes Bob’s accessible information when Alice selects that specific configuration, as fundamentally bounded by Holevo's theorem~\cite{Holevo1973seminal}.

The classical correlation in the state $\rho_{AB}$ as defined by Henderson and Vedral is then the maximal Holevo quantity obtained by optimizing over all local projective measurements on Alice's system~\cite{HV01}:
\begin{equation}
C_{1}(\rho_{AB}) \equiv \max_{\{\left|a_{i}\right\rangle _{A}\}}\chi\big\{\rho_{AB}\big|\{\left|a_{i}\right\rangle _{A}\}\big\}.\label{defC1}
\end{equation}
The optimal configurations optimizing Eq.~\eqref{defC1} are designated as $C_1$-bases, denoted by $\left\{ \left|\mathcal{A}_{i}^{1}\right\rangle _{A}\right\} $~\cite{WMC+14}. Crucially, this optimized classical information quantity $C_1(\rho_{AB})$ is equivalent to the Henderson--Vedral classical correlation measure $J^{\rightarrow}(\rho_{AB})$ utilized to split the cooperative work of assistance in Eq.~\eqref{workass4}.

A defining feature of genuine quantum correlation is that projecting onto an alternative basis $\left\{ \left|a_{j}^{2}\right\rangle _{A}\right\}$ mutually unbiased to the $C_1$-basis—where $\left|\left\langle \mathcal{A}_{i}^{1}|a_{j}^{2}\right\rangle \right|=1/\sqrt{d_{A}}$—leaves nonvanishing residual Holevo information \cite{WMC+14}. The maximum accessible information under this second MUB setup defines the residual quantum correlation metric $Q_2$~\cite{WMC+14}:
\begin{equation}
Q_{2}(\rho_{AB}) \equiv \max_{\{\left|\mathcal{A}_{i}^{1}\right\rangle _{A}\}} \max_{\{\left|a_{j}^{2}\right\rangle _{A}\}}\chi\big\{\rho_{AB}\big|\{\left|a_{j}^{2}\right\rangle _{A}\}\big\} . \label{defQ2}
\end{equation}
Iterating this optimization structure into a third MUB configuration yields the residual correlation $Q_3$~\cite{GW14}:
\begin{equation}
Q_{3}(\rho_{AB}) \equiv \max_{\{\left|\mathcal{A}_{i}^{1}\right\rangle _{A}\}} \max_{\{\left|\mathcal{A}_{j}^{2}\right\rangle _{A}\}} \max_{\{\left|a_{k}^{3}\right\rangle _{A}\}}\chi\big\{\rho_{AB}\big|\{\left|a_{k}^{3}\right\rangle _{A}\}\big\},
\label{defQ3}
\end{equation}
where $\{\left|a_{k}^{3}\right\rangle _{A}\}$ is simultaneously unbiased to both the $C_1$ and $Q_2$ bases \cite{GW14}.

Alternatively, the maximum simultaneous correlation distributed across arbitrary MUB sets can be quantified natively~\cite{GW14}. Let $\Delta$ represent the collection of all two-MUB pairs~\cite{GW14}. The simultaneous correlation metric $C_2$ is defined as~\cite{GW14}:
\begin{equation}
C_2(\rho_{AB}) \equiv \max_{(\Pi^a_{1},\Pi^a_{2})\in\Delta} \min\big\{ \chi\{\rho_{AB}|\{\Pi^a_{1}\}\},\chi\{\rho_{AB}|\{\Pi^a_{2}\}\}\big\} . \label{C2def}
\end{equation}
We have $Q_2 \le C_2$ because $C_2$ is nonzero for any nonproduct state, whereas $Q_2>0$ requires global coherence~\cite{Jeb+26_GlobalCoherence} in quantum discord~\cite{OZ01, HV01}.  
Extending this framework to triads of MUBs $\Lambda_A$ within Alice's Hilbert space defines the three-setting simultaneous correlation metric $C_3$ \cite{GW14}:
\begin{align} \label{C3def}
    C_3 (\rho_{AB})&= \max_{\{\Pi_k^A\}_1^3 \in \Lambda_{A}} \min \Big[ \chi(\rho_{AB}|\{\Pi_1^A\}), \nonumber \\ 
    &\quad \chi(\rho_{AB}|\{\Pi_2^A\}), \chi(\rho_{AB}|\{\Pi_3^A\}) \Big].
\end{align}
We have $Q_3\le C_3$. To clarify the distinct optimization regimes, a fundamental structural difference exists between these two sets of metrics. For the residual correlation parameters $Q_2$ and $Q_3$, the primary measurement configuration is strictly restricted to be the optimal $C_1$-basis maximizing the classical correlation~\cite{WMC+14, GW14}. The remaining configurations are then optimized sequentially over bases that are mutually unbiased to this fixed baseline~\cite{WMC+14, GW14}. In contrast, for the simultaneous correlation metrics $C_2$ and $C_3$, no single basis is anchored to the classical optimum; instead, the optimization is performed globally over the entire space of valid MUB pairs or triads to maximize their concurrent minimum shared information~\cite{GW14}. Here, the minimum shared information refers to the smallest Holevo quantity obtained among the chosen settings; maximizing this lower bound ensures that nonclassical correlations are simultaneously present and captured maximally across all complementary bases, rather than being localized within a single preferred basis configuration.

\subsection{Quantum steering vs SCMUB}
In an operational steering scenario~\cite{WJD07}, Alice performs a set of black-box local measurements labeled by $x$ with outcomes $a$, which prepares a conditional state ensemble on Bob's side. This scenario is formally characterized by an assemblage $\sigma_{a|x} = p(a|x)\rho_{a|x}$, or equivalently by the joint correlation box $P(ab|A_x B_y) = \Tr[(M_{a|x} \otimes N_{b|y})\rho_{AB}]$, where $M_{a|x}$ and $N_{b|y}$ denote the local POVM elements. EPR steering is certified if and only if the assemblage cannot be reproduced by an LHS model; such a certification strictly requires the presence of entanglement~\cite{WJD07,PW15}.

For a two-qubit state $\zeta$ written in the standard representation~\cite{Luo08}:
\begin{align}
\zeta &= \frac{1}{4} \Big( \openone \otimes \openone + \vec{a} \cdot \vec{\sigma} \otimes \openone + \openone \otimes \vec{b} \cdot \vec{\sigma} \nonumber \\
&\quad + \sum^3_{i=1} t_i \sigma_i \otimes \sigma_i \Big), \label{can02q}
\end{align}
the steerability can be quantified via the maximal violation of the linear Cavalcanti-Jones-Wiseman-Reid (CJWR) steering inequalities~\cite{CJW+09,CA16}:
\begin{align}
S_n(\rho) &= \max\left\{0,\frac{F_n(\rho)-1}{F^{\mathrm{max}}_n-1}\right\}, \label{SQ}
\end{align}
where $F_n(\rho) = \max_{\mu} \frac{1}{\sqrt{n}} \left|\sum^n_{k=1} \braket{A_k\otimes B_k}\right|$~\cite{CJW+09,CA16}. For two- and three-setting setups, this yields:
\begin{align}
F_2(\zeta) &= \sqrt{t^2-t^2_{\min}}, \\
F_3(\zeta) &= t,
\end{align}
where $t$ is the norm of the correlation vector $\vec{t}$ and $t_{\min}=\min\{|t_1|,|t_2|,|t_3|\}$~\cite{CA16}.

For the two-qubit Bell-diagonal state family $\tau = \frac{1}{4}(\openone \otimes \openone +\sum_i t_i \sigma_i \otimes \sigma_i)$, the simultaneous MUB correlation metrics have closed expressions~\cite{GW14}:
\begin{align} 
C_2(\tau) &= 1-h\left(\frac{1+\sqrt{(t^2-t^2_{\min})/2}}{2}\right), \label{C2BD} \\
C_3(\tau) &= 1-h\left(\frac{1+t/\sqrt{3}}{2}\right), \label{C3BD}
\end{align}
where $h(x)$ is the binary entropy function \cite{GW14}. These metrics map directly to the steering parameters via monotonic relationships \cite{JKK+18}:
\begin{align}
C_2(\tau) &= 1-h\left(\frac{1+F_2(\tau)/\sqrt{2}}{2}\right), \label{C2F2} \\
C_3(\tau) &= 1-h\left(\frac{1+F_3(\tau)/\sqrt{3}}{2}\right). \label{C3BDF3}
\end{align}
Equations~\eqref{C2F2} and \eqref{C3BDF3} reveal that $C_2$ and $C_3$ serve as explicit quantitative resources for quantum steering within Bell-diagonal states~\cite{JKK+18}.

\subsection{1SSDI steering vs SCMUB}
When an assemblage admits a standard LHS description, nonclassicality can still be verified under 1SSDI frameworks by restricting the untrusted party to a trusted dimension ($d_A=2$)~\cite{Jebarathinam2023Asymmetric,Jeb+25,Jeb+26}. Within this paradigm, 1SSDI steering is confirmed if for any measurements on Bob's side, the box $P(ab|xy) = \{p(ab|xy)\}$ produced from the given assemblage   cannot be simulated by any local hidden variable-local hidden state (LHV-LHS) model, restricted to a hidden state space dimension $d_\lambda \le d_A$~\cite{DBD+18}. This dimensional constraint operates as a direct bridge to the LHS separability framework of Fadel and Gessner~\cite{fadel2022entanglement}, which demonstrates that while LHS models describe assemblages that appear classical under unconstrained scenarios, the underlying hidden ensembles themselves can harbor nontrivial  entanglement structures. 

Operationally probing this hidden domain, the underlying capacity of such dimension-restricted unsteerable assemblages was originally characterized and investigated using the definition of \textit{superunsteerability}~\cite{DBD+18,JDK+19,jebarathinam2023}. This resource capability was subsequently formalized under the framework of asymmetric 1SSDI steerability in Ref.~\cite{Jebarathinam2023Asymmetric,Jeb+25}. By explicitly bounding the Hilbert space dimension of the untrusted observer, the 1SSDI framework maps the conceptual boundaries of the LHS entanglement hierarchy introduced in Ref.~\cite{fadel2022entanglement} onto an active certification scheme, demonstrating that imposing a dimensional trust can activate and witness the nonclassical utility of states that otherwise exhibit standard LHS descriptions~\cite{datta2008quantum,Jeb+25,Jeb+26,Jeb+26_GlobalCoherence}.

The operational advantage of 1SSDI steering resource can be captured using the 1SSDI-restricted Schr\"{o}dinger strength $S_{\text{1SSDI}_n}(\rho)$, which optimizes the steerable fraction of an assemblage over all local measurement configurations in the given $n$-setting scenario~\cite{JDK+19,Jeb+25,Jeb+26_GlobalCoherence}.
The standard Schr\"odinger strength refers to the operational framework introduced in~\cite{DDJ+18}, which quantifies the operational cost of EPR steering in the context of extremal boxes.
For two-qubit Bell-diagonal states, the two- and three-setting 1SSDI strengths collapse to $S_{\text{1SSDI}_2}(\tau)=|t_2| $ and $S_{\text{1SSDI}_3}(\tau)=|t_3|$ \cite{JDK+19}, which are bound monotonically to the information-theoretic complementary measures $Q_2$ and $Q_3$ via~\cite{WMC+14,GW14,Jeb+25}:
\begin{align}
Q_2(\tau) &= 1-h\left(\frac{1+S_{\text{1SSDI}_2}(\tau)}{2}\right), \label{Q2SU2} \\
Q_3(\tau) &= 1-h\left(\frac{1+S_{\text{1SSDI}_3}(\tau)}{2}\right). \label{Q3BDS3}
\end{align}
Thus, the complementary residual parameters $Q_2$ and $Q_3$ directly quantify the capacity of Bell-diagonal systems to implement 1SSDI steering protocols~\cite{Jeb+25}.
Having established the  monotonic relationships between dimension-bounded steering strengths and information-theoretic MUB measures, we now leverage this foundation to construct an operational framework for quantum work extraction.

\section{Work of Assistance under Unconstrained and Dimension-Bounded Steering Regimes}
\begin{figure}[htbp]
    \centering
    \includegraphics[width=0.72\textwidth]{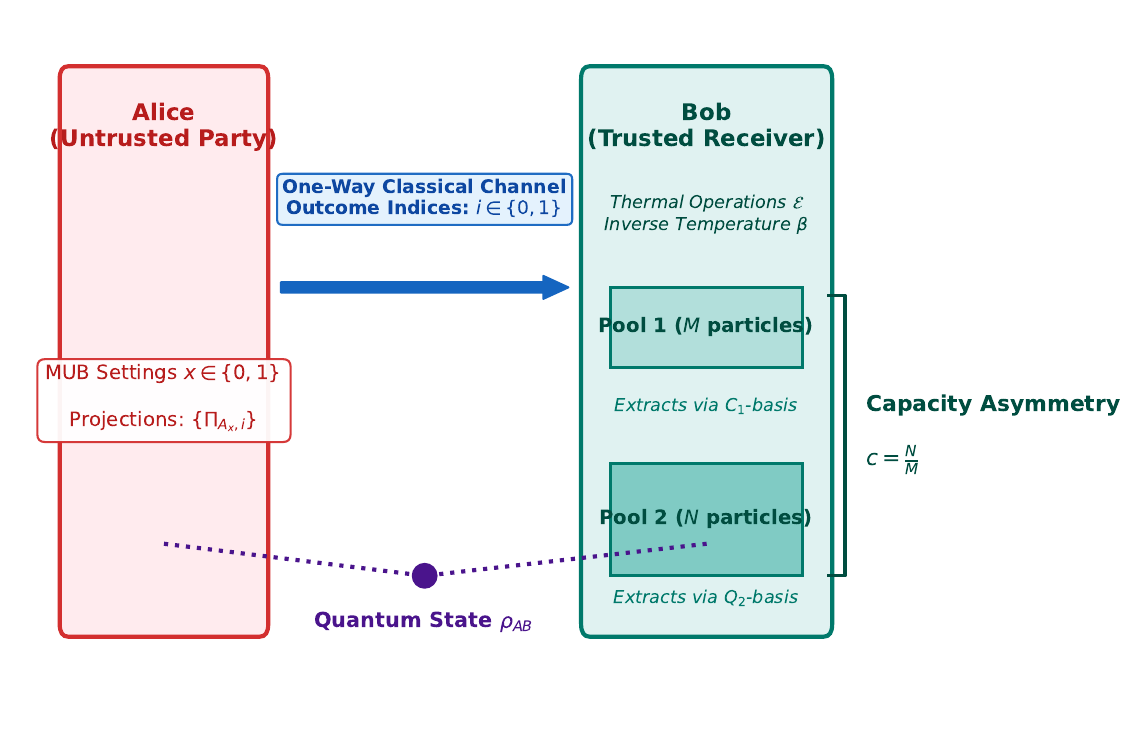}
    \caption{Schematic layout of the asymmetric assisted work distillation framework under 
    two-setting complementary measurement configurations~\cite{MLA19}. An unconstrained 
    observer (Alice) and a localized receiver (Bob) share a bipartite quantum state $\rho_{AB}$ 
    across a spatial divide. Bob's local storage system is physically divided into separate 
    particle ensembles or storage pools of sizes $M$ and $N$, respectively. To act as a 
    microscale quantum Maxwell's demon, Alice performs local projections within a set of 
    MUBs and transmits her raw measurement index $i$ over a 
    one-way classical channel. Bob then applies suitable local background TO at inverse temperature $\beta$ to extract work from Pool $1$ via the optimal 
    classical $C_1$-basis and from Pool $2$ via the complementary $Q_2$-correlation setting. 
    The relative capacity distribution between these prepared ensembles defines the core 
    thermodynamic asymmetry parameter $c = N/M$, which directly governs the dynamic collapse 
    of the dimension-bounded classical validation thresholds.}
    \label{fig:thermodynamic_demon_setup}
\end{figure}

Extending the scenario of Morris \etal~\cite{MLA19} for enhancement in the distillable work, we consider a microscale thermodynamic setup where Bob's localized system is physically divided into separate particle ensembles or storage pools, each designated for work extraction under a distinct measurement setting, as visually illustrated in the schematic layout of Fig.~\ref{fig:thermodynamic_demon_setup}. As illustrated in the protocol architecture, an unconstrained observer  (Alice) performs local projections within a chosen set of 
MUBs and transmits her measurement outcome $i$ over a 
one-way classical channel. To act as a thermodynamic quantum Maxwell's demon, Bob 
leverages this classical indicator to apply local TO, optimally distilling 
work from each conditionally prepared pool. The relative operational capacity weights 
between these storage pools define our core thermodynamic asymmetry parameters. For the 
two-setting scenario, this is formalized as the capacity ratio $c = N/M$ between Pool 1 
($M$ particles, optimized for work extraction via the classical $C_1$-basis) and Pool 2 
($N$ particles, optimized for extraction via the complementary $Q_2$-basis setting). We 
then extend this configuration to three-setting scenarios involving three complementary 
bases by introducing a third storage pool of capacity $O$, establishing the joint capacity 
distribution vector $(c, d)$, where $d = O/M$. As we show in the following sections, this 
unified formulation outlines a powerful nonclassical crossover behavior where highly asymmetric 
ensemble distributions collapse the dimension-bounded classical validation thresholds, 
unlocking the hidden thermodynamic utility of unsteerable states through information-theoretic 
measures of SCMUB~\cite{WMC+14,GW14}.
\subsection{Work of assistance in two-setting steering scenario}
Let us now consider the scenario of Morris \etal \cite{MLA19} for enhancement in the distillable work via two different quantum measurements on Alice's side, labeled by $x \in \{0, 1\}$. In this framework, Bob's localized storage system is physically divided into two separate particle ensembles or pools containing $M$ and $N$ number of particles, where the relative operational capacity ratio is given by $c = N/M$. To unlock a thermodynamic quantum advantage, for these two pools, Bob asks Alice to prepare two different ensembles such that he can optimally distill work by applying two different TO on these two ensembles prepared by Alice.

In order to exploit quantum correlation to provide enhancement in the distillable work, let Alice perform two different quantum measurements denoted $A_0$ and $A_1$ in MUBs. We now ask the question whether the measures of SCMUB can be exploited for quantifying the quantum advantage for local work extraction from the bipartite quantum systems in this context. To this end, let us consider the average work of assistance defined as:
\begin{equation}
  AW^{B|A}_a := \frac{W_a^{B|A_0}(\rho_{AB}) + c W_a^{B|A_1}(\rho_{AB})}{1+c}, \label{bls19}
\end{equation}
where the work of assistance due to the measurements $\{\Pi_{A_x,i}\}$ is given by:
\begin{align}
  W_a^{B|A_0}(\rho_{AB})&=\frac{1}{\beta}\sum_i p_i S(\tilde{\rho}_{B, i|A_0}||\gamma_B) \nonumber \\
  &=\frac{1}{\beta}\left(S(\rho_B||\gamma_B)+ \chi\big\{\rho_{AB}\big|\{\Pi_{A_0,i}\}\big\} \right), \label{eq:wa_setting0}
\end{align}
where $\chi(\rho_{AB}\big|\{\Pi_{A_0,i}\})=S(\rho_B)- \sum_i p_i S(\tilde{\rho}_{B,i|A_0})$ and 
\begin{align}
  W_a^{B|A_1}(\rho_{AB})&=\frac{1}{\beta}\sum_i p_i S(\tilde{\rho}_{B, i|A_1}||\gamma_B) \nonumber \\
  &=\frac{1}{\beta}\left(S(\rho_B||\gamma_B)+ \chi\big\{\rho_{AB}\big|\{\Pi_{A_1,i}\}\big\} \right), \label{eq:wa_setting1}
\end{align}
where $\chi\big\{\rho_{AB}\big|\{\Pi_{A_1,i}\}\big\}=S(\rho_B)- \sum_i p_i S(\tilde{\rho}_{B,i|A_1})$, with $\tilde{\rho}_{B, i|A_x}$ being the conditional states due to the measurements $\{\Pi_{A_x,i}\}$.
The above average work of assistance is conceptually similar to the quantity considered in Ref.~\cite{BLS19}, where average work extraction in the presence of two different quantum measurements was introduced to derive steering inequalities (see also~\cite{BiswasWork2025, BiswasCooling2026}). Likewise, in this work, we derive a steering inequality directly in terms of the average work of assistance \eqref{bls19}.

\subsubsection{Steering as a resource}

To certify a genuine quantum advantage in this two-setting framework, we evaluate the average work extraction given by Eq.~\eqref{bls19} against the classical benchmark established by the LHS framework~\cite{WJD07}. That is, we derive a steering inequality of the form $AW^{B|A}_a \le W_{LHS}$,
where $W_{LHS}$ is the LHS bound on $AW^{B|A}_a$. Thus, in our approach to the steering inequality in this asymmetric thermodynamic context,  the operational quantification of steerability  is more naturally captured through the optimization of the steerable weight of the prepared  assemblage~\cite{Skrzypczyk2014steerable,Cavalcanti2017review}. There exists a well-established dual relationship between the steerable weight of an  assemblage and the maximal violation of an optimized linear steering inequality~\cite{Skrzypczyk2014steerable}. This implies that the optimal measurements for both aforementioned steering witnesses are the same.

To derive our steering inequality, we make use of the properties of the family of Bell-diagonal two-qubit  states.
It has been demonstrated that this family of two-qubit  states optimizes the violation of linear steering inequalities under cooperative local filtering operations~\cite{Ducuara2024hidden}. For the Bell-diagonal states, the MUBs that achieve the maximal simultaneous correlation $C_2(\tau)$ correspond precisely to the unique spatial Bloch directions $\hat{n}_{\pm} = (\frac{1}{\sqrt{2}}, \pm\frac{1}{\sqrt{2}}, 0)$ which optimize the underlying linear steering parameter $F_2$~\cite{CA16,JKK+18}. When Alice performs local qubit projective measurements within these optimal configurations to assist Bob, Bob is not assumed to perform fixed  measurements to check steerability for our thermodynamic demon. Instead, Bob employs 
suitable background TO to extract work optimally from each conditionally prepared ensemble, so that the work of assistance $W_a^{B|A_x}(\tau)$ for each setting simplifies uniformly to $W_a^{B|A_x}(\tau) = \frac{1}{\beta}\left(S(\rho_B||\gamma_B) + C_2(\tau)\right)$. Consequently, the weighted average work of assistance defined in Eq.~\eqref{bls19} evaluates cleanly to:
\begin{equation} \label{C_2Wa_revised} 
AW^{B|A}_a(\tau) = \frac{1}{\beta}\left(S(\rho_B \Vert \gamma_B) + C_2 (\tau) \right).
\end{equation}
Using this explicit functional link alongside the monotonic relationship established in Eq.~\eqref{C2F2} between $F_2(\tau)$ and $C_2(\tau)$, the average work of assistance for any classical LHS ensemble arising from an arbitrary shared state $\rho_{AB}$ is strictly upper-bounded by:
\begin{equation}\label{wasteerineq_revised} 
AW^{B|A}_a(\rho_{AB}) \le \frac{1}{\beta}\left(S(\rho_B||\gamma_B) + 1-h\left(\frac{1+1/\sqrt{2}}{2}\right) \right). 
\end{equation}
This bound  follows because $F_2 \le 1$  for any valid LHS model~\cite{CJW+09}, and the linear steering parameter $F_2(\rho_{AB})$ is lower-bounded by the canonical Bell-diagonal state obtained from $\rho_{AB}$ via optimal local filters~\cite{Ducuara2024hidden}.

Any experimental observation utilizing any given shared quantum state and local measurements on Alice's side aligned along symmetric mutually unbiased directions that optimize simultaneous correlations ($C_2$)—yielding a value of $AW^{B|A}_a(\rho_{AB})$ exceeding the right-hand side of Eq.~(\ref{wasteerineq_revised})—certifies quantum steering.
 Because the LHS bound of the inequality (\ref{wasteerineq_revised})   holds regardless of the dimensionality or internal mechanism of the untrusted party (Alice), this violation serves as a  one-sided device-independent steering inequality stated in terms of micro-thermodynamic work distillation. 

Furthermore, using Eq.~(\ref{C_2Wa_revised}), it follows that the optimal violation of the steering inequality (\ref{wasteerineq_revised}) by using a Bell-diagonal state provides an operational characterization of quantum steering characterized by the amount of simultaneous correlation captured via $C_2$.  Thus, the thermodynamic quantum enhancement using quantum steering of Bell-diagonal states is explicitly proportional to the amount of simultaneous correlation $C_2$ captured across the complementary bases, demonstrating how quantumness related to  steering acts as a direct fuel for quantum Maxwell's demons.

\subsubsection{1SSDI steering as a resource}
Next, we investigate how the average work of assistance in Eq.~(\ref{bls19}) in the 1SSDI steering framework provides nonclassical thermodynamic utility. To this end, similar to the steering inequality in Eq.~(\ref{wasteerineq_revised}), we derive a 1SSDI steering inequality.
To derive such an inequality, we note that for any classical-quantum  (CQ) two-qubit state which does not exhibit 1SSDI steering~\cite{Jebarathinam2023Asymmetric,Jeb+25}, the separability threshold $F_2 \le 1/2$ holds~\cite{CHR+11}. Using this inequality to witness the 1SSDI steering,  we obtain the following limitation on the average work of assistance:
\begin{equation}\label{waunsteerineq}
 AW^{B|A}_a(\rho_{AB}) \le \frac{1}{\beta}\left(S(\rho_B||\gamma_B) + 1-h\left(\frac{1+1/2}{2}\right) \right),
\end{equation}
under the constraint of LHS models in the 1SSDI context. To obtain the above inequality, we have also made use of the functional link in Eq.~(\ref{C_2Wa_revised}) alongside the monotonic relationship established in Eq.~\eqref{C2F2} between $F_2(\tau)$ and $C_2(\tau)$. In the above derivation of the inequality, we have employed the bound achieved by the set of CQ states in the 1SSDI context, which is inspired by the standard steering context~\cite{Cavalcanti2017review,Moroder2016SteeringMaps} in which case every LHS model can be reproduced by a CQ state of the form:   \begin{align}\label{rhoCQ_def}
\rho_{CQ} = \sum_{\lambda} p_\lambda \proj{\lambda}_A \otimes \rho_{B|\lambda},
\end{align}
where $p_\lambda$ represents the probability distribution over the hidden variables, and the set of local states $\{\ket{\lambda}_A\}$ form an orthonormal basis and the states $\rho_{B|\lambda}$ are an arbitrary local quantum state.

It can be immediately inferred that a violation of Eq.~(\ref{waunsteerineq}) confirms that certain unsteerable entangled states, which have $F_2>1/2$, can be leveraged for quantum work distillation in the 1SSDI scenario. Comparing Eqs.~(\ref{wasteerineq_revised}) and (\ref{waunsteerineq}), it follows that certain states which possess 1SSDI steering—and cannot be used to provide a quantum advantage in the unconstrained setup where Alice is fully device independent—can successfully demonstrate a quantum advantage when the dimension of Alice is bounded. In this case, $C_2(\tau)$ directly quantifies the quantum advantage using a Bell-diagonal state as in the case of inequality (\ref{wasteerineq_revised}).

 Note that the parameter $c = N/M$ does not play any role in addressing quantum work enhancement in the context of the inequalities that we derived before. To explore its role in quantum work, we adopt an  adaptive scheme depicted in Fig.~\ref{fig:thermodynamic_demon_setup}. Since $Q_2$ uniquely quantifies this 1SSDI resource capability~\cite{Jeb+25}, deploying a $C_1$-basis projection for Alice's first measurement optimizes the average work expression by cleanly partitioning the available correlations into classical ($C_1$) and residual quantum ($Q_2$) components. From Eq.~(\ref{workass4}), we have $W_a^{B|A_0}(\rho_{AB}) = \frac{1}{\beta}\left(S(\rho_B||\gamma_B) + C_1(\rho_{AB})\right)$, since $J^{\rightarrow}(\rho_{AB}) = C_1(\rho_{AB})$ under a $C_1$-basis projection. Moreover, since the second basis is chosen to be mutually unbiased to the $C_1$-basis, the second measurement yields $W_a^{B|A_1}(\rho_{AB}) = \frac{1}{\beta}\left(S(\rho_B||\gamma_B) + Q_2(\rho_{AB})\right)$. Hence, the average work mapping from Eq.~(\ref{bls19}) can be written as:
\begin{align}\label{AW_1SSDI_two_setting}
     AW^{B|A}_a(\rho_{AB}) = \frac{1}{\beta}\left(S(\rho_B||\gamma_B) + \frac{ C_1(\rho_{AB}) + c Q_2(\rho_{AB}) }{1+c} \right).   
\end{align}
For configurations devoid of 1SSDI steering, this average work is strictly bounded by:
\begin{align}\label{WaQ2steer}
   AW^{B|A}_a(\rho_{AB}) \le \frac{1}{\beta}\left(S(\rho_B||\gamma_B) + \frac{1}{1+c} \right).
\end{align}
Equation~\eqref{WaQ2steer} represents a valid, dynamically scaling 1SSDI steering inequality for dimension-bounded setups. The formal mathematical proof of this boundary condition is derived explicitly  in~\ref{app:derivation}. 

It is crucial to clarify the distinct operational and foundational regimes captured by the 1SSDI steering witnesses presented in Eq.~(\ref{waunsteerineq}) and Eq.~(\ref{WaQ2steer}). While both inequalities successfully certify a dimension-bounded quantum enhancement, Eq.~(\ref{waunsteerineq}) relies on a static symmetric setting designed to map the simultaneous correlation $C_2$ uniformly across both measurement bases. In contrast, Eq.~(\ref{WaQ2steer}) assumes an adaptive configuration where Alice's first observable is dynamically aligned with the state's optimal $C_1$-basis, explicitly parsing the remaining nonclassical utility through the residual correlation $Q_2$. 

Furthermore, unlike the constant classical threshold of Eq.~\eqref{waunsteerineq}, which remains fixed at $1-h(3/4) \approx 0.1887$, the upper bound of Eq.~\eqref{WaQ2steer} scales dynamically as $\frac{1}{1+c}$. Equating these two boundaries yields the exact threshold capacity ratio $c^* = \frac{h(3/4)}{1-h(3/4)} \approx 4.30$. We explicitly designate the parameter region where $c < c^* \approx 4.30$ as the \textit{symmetric domain}. Within this regime, the capacity of the storage pools is sufficiently balanced so that the static, dimension-bounded classical threshold dominates the verification landscape. This dominance arises because the directed asymmetric boundary remains higher than the static baseline (i.e., $\frac{1}{1+c} > 0.1887$), rendering the constant symmetric bound ($\approx 0.1887$) the stricter and more physically relevant benchmark for certifying a nonclassical thermodynamic advantage. Conversely, in highly asymmetric particle regimes where $c > 4.30$, this dynamic boundary collapses significantly below the static 1SSDI threshold. This structural collapse renders Eq.~\eqref{WaQ2steer} a noticeably more sensitive witness in this asymmetric domain, capable of activating and certifying the hidden thermodynamic utility of weakly correlated entangled states that are otherwise completely invisible to symmetric setups.

Let us now consider the Bell-diagonal states $\tau$ with $|t_1| \ge |t_2| \ge |t_3|$. For this class of two-qubit states, $C_1(\tau)=1-h\left(\frac{1+|t_1|}{2}\right)$ and $Q_2(\tau)=1-h\left(\frac{1+|t_2|}{2}\right)$. Therefore, we obtain:
\begin{equation}\label{Q_2BDwa_clean}
\begin{split}
   AW^{B|A}_a (\tau) = \, & \frac{1}{\beta} S(\rho_B||\gamma_B) \\
   & + \frac{1}{\beta(1+c)} \left[1 - h\left(\frac{1+|t_1|}{2}\right)\right] \\
   & + \frac{c}{\beta(1+c)} \left[1 - h\left(\frac{1+|t_2|}{2}\right)\right].
\end{split}
\end{equation}
From this expression, it follows that the Bell-diagonal states demonstrate a quantum advantage with respect to the inequality given by Eq.~\eqref{WaQ2steer} if:
\begin{align}\label{two_setting_1SSDI_condition}
C_1+cQ_2=1-h\left(\frac{1+|t_1|}{2}\right)+c\left[1- h\left(\frac{1+|t_2|}{2}\right)\right]>1.
\end{align}
The enhancement in distillable work is proportional to the amount of SCMUB as $(C_1+cQ_2)/(1+c)$, implying that the dimension-bounded quantum advantage is simultaneously quantified by both $C_1$ and $Q_2$ correlations~\cite{WMC+14,GW14,Jeb+25}.

\begin{figure}[!t]
    \centering
    \includegraphics[width=0.72\textwidth]{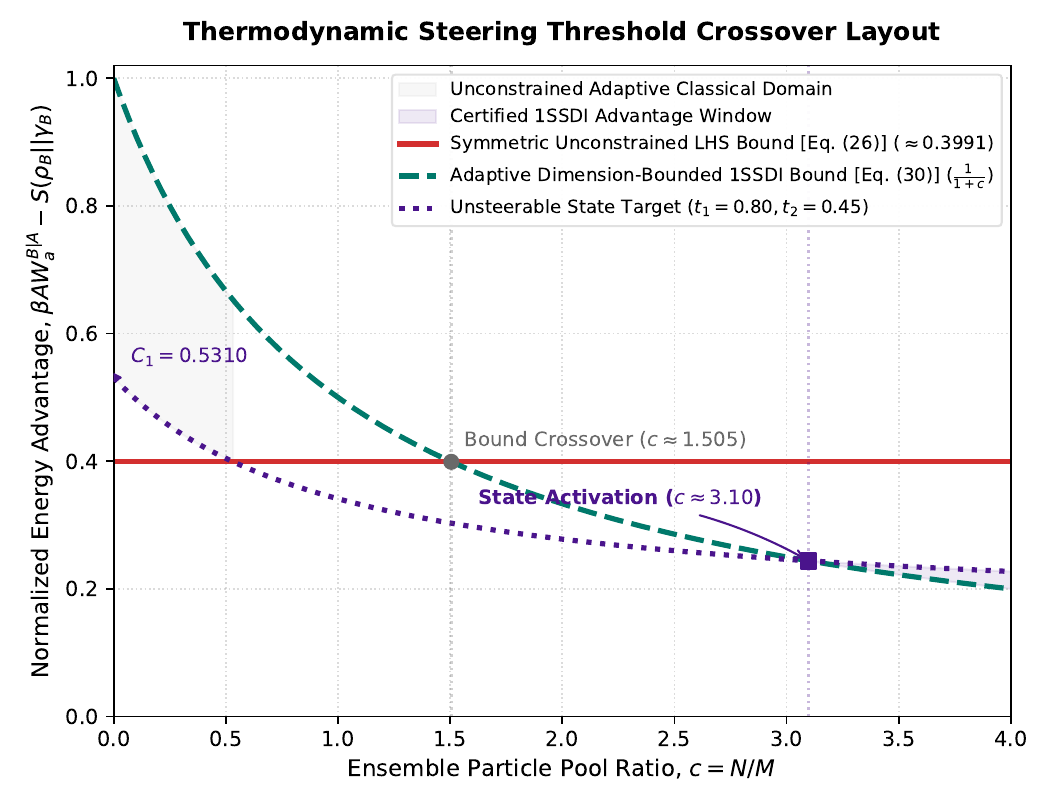}
    \caption{The normalized average work enhancement as a function of the particle ensemble ratio $c = N/M$. The solid horizontal line marks the symmetric unconstrained LHS threshold ($\approx 0.3991$), while the dashed curve outlines the adaptive dimension-bounded 1SSDI threshold $\frac{1}{1+c}$. At the far left ($c \to 0$), the target state profile line ($t_1=0.80, t_2=0.45$) begins at its optimal classical correlation coordinate ($C_1 \approx 0.5310$), placing it above the symmetric baseline within the uncertified unconstrained adaptive domain. Note that the static bound of Eq.~(\ref{wasteerineq_revised}) ($\approx 0.3991$) applies exclusively when Alice measures in the symmetric $C_2$-optimal MUBs (for which this target state yields $C_2 \approx 0.3125 < 0.3991$), whereas the adaptive $C_1/Q_2$ protocol plotted here must be benchmarked against the dynamically scaling bound $\frac{1}{1+c}$ of Eq.~(\ref{WaQ2steer}). As the system moves into highly asymmetric regimes ($c > 3.10$), the state line comfortably violates this collapsing 1SSDI boundary while remaining strictly below the unconstrained horizontal baseline, certifying its hidden thermodynamic utility.}
    \label{fig:threshold_crossover}
\end{figure}

This functional dependency and its resulting thermodynamic implications are explicitly illustrated in Fig.~\ref{fig:threshold_crossover}. Here, we monitor the normalized work advantage as a function of the ensemble configuration ratio $c$. The flat crimson boundary represents the symmetric unconstrained LHS threshold, which remains static at $\approx 0.3991$. Conversely, the dimension-bounded 1SSDI classical boundary drops as $\frac{1}{1+c}$, establishing an exact geometric intersection with the unconstrained horizontal at the crossover point $c \approx 1.505$. 

By tracking an unsteerable Bell-diagonal state parameterized by $t_1 = 0.80$ and $t_2 = 0.45$, the plot captures a powerful nonclassical feature. At the far left ($c \to 0$), the target state profile begins at its optimal classical correlation coordinate ($C_1 \approx 0.5310$), placing it above the symmetric baseline within the unconstrained adaptive classical domain. As we migrate through the intermediate pool regime ($c \le 3.10$), the state profile remains safely below the active classical bounds. However, once the ensemble pool asymmetry becomes highly pronounced ($c > 3.10$), the dynamic 1SSDI boundary collapses strictly below the state's available correlation capability. Consequently, the target state comfortably breaches the 1SSDI boundary while remaining strictly below the unconstrained horizontal baseline, providing a rigorous proof that unsteerable states house a hidden thermodynamic utility that can be unlocked and verified by asserting dimensional trust.

To anchor this specific state configuration to a concrete physical layout, we observe that any Bell-diagonal two-qubit state can be decomposed as a convex combination of the four maximally entangled Bell states, $\tau = \sum_{ab} \lambda_{ab} \vert\beta_{ab}\rangle\langle\beta_{ab}\vert$, with eigenvalues $\lambda_{\Phi^{\pm}} = (1 \pm t_1 \mp t_2 + t_3)/4$ and $\lambda_{\Psi^{\pm}} = (1 \pm t_1 \pm t_2 - t_3)/4$. Choosing the physical unsteerable parameters $t_1 = 0.80$, $t_2 = 0.45$, and $t_3 = -0.25$ (so that $|t_1| = 0.80 \ge |t_2| = 0.45 \ge |t_3| = 0.25$) yields non-negative eigenvalues $\lambda_{\Phi^+} = 0.625$, $\lambda_{\Psi^+} = 0.275$, $\lambda_{\Psi^-} = 0.100$, and $\lambda_{\Phi^-} = 0$. This configuration corresponds to an exact operational mixture composed of $62.5\%$ of the Bell state $\vert\Phi^+\rangle = \frac{1}{\sqrt{2}}(\vert 00 \rangle + \vert 11 \rangle)$, $27.5\%$ of $\vert\Psi^+\rangle = \frac{1}{\sqrt{2}}(\vert 01 \rangle + \vert 10 \rangle)$, and $10.0\%$ of $\vert\Psi^-\rangle = \frac{1}{\sqrt{2}}(\vert 01 \rangle - \vert 10 \rangle)$. While statistical mixing reduces the unconstrained steering metrics to $F_2 \approx 0.9179 \le 1$ (and $F_3 \approx 0.9513 \le 1$), masking its nonclassicality under unconstrained scenarios, the dimensionally bounded trust framework successfully filters out this noise background to extract genuine thermodynamic quantum utility from the underlying entangled components.

\subsection{Work of assistance in three-setting steering scenario}
Let us now consider the scenario of Morris \etal~\cite{MLA19} for enhancement in the distillable work with three different quantum measurements on Alice's side labelled by $x \in \{0,1,2\}$. In such a scenario, the ensemble of Bob's particles is divided into three sets of particles having $M$, $N$, and $O$ number of particles, where the ratios between the particle numbers across these three sets are given by $c=N/M$ and $d=O/M$. For these three sets of particles, Bob asks Alice to prepare three different ensembles such that he can optimally distill work by applying three distinct local background TO on these three ensembles respectively.

In order to exploit genuine quantum correlation in a three-setting scenario to provide enhancement in the distillable work, let Alice perform three different quantum measurements denoted $A_0$, $A_1$, and $A_2$ in MUBs. We now ask the question whether the measures of SCMUB defined with respect to the three MUBs can be exploited for quantifying quantum advantage for local work extraction from the bipartite quantum systems in this context.

\subsubsection{Steering as a resource}

Consistently extending this dimension-free paradigm, the operational verification of nonclassicality within the three-setting configuration avoids any restrictive assumptions regarding fixed measurement devices on the trusted receiver's side. As in the two-setting setup, the  nonclassical capacity of our thermodynamic systems is rigorously captured through the optimization of the steerable weight over a prepared three-measurement quantum assemblage~\cite{Skrzypczyk2014steerable,Cavalcanti2017review}. Exploiting the dual relationship that maps the maximum steerable fraction directly onto the supremum of an optimized linear steering inequality, mandating maximal steering verification under three settings dictates the exact spatial configurations of Alice's local observables. For the family of two-qubit Bell-diagonal states, the three MUBs that fully optimize this steerable weight match perfectly with the MUBs that yield the maximal simultaneous correlation $C_3(\tau)$ distributed concurrently across three complementary axes~\cite{GW14}.

When Alice performs local projective measurements within this optimal triad of bases to assist Bob, Bob remains unconstrained to any fixed measurement setups. Rather, Bob employs a set of local background TO to extract work optimally from each conditionally steered particle pool. Under this optimized configuration, the work distilled from each of the three individual measurement configurations simplifies uniformly to a value bounded by the three-setting simultaneous correlation, yielding $W_a^{B|A_x}(\tau) = \frac{1}{\beta}\left(S(\rho_B||\gamma_B) + C_3(\tau)\right)$. Consequently, the weighted average work of assistance defined across the three distinct sets of storage systems collapses directly to:
\begin{equation} \label{C_3Wa_revised}
    AW^{B|A}_a(\tau) = \frac{1}{\beta}\left(S(\rho_B||\gamma_B) + C_3 (\tau) \right).
\end{equation}
Using this explicit functional link alongside the monotonic relationship established in Eq.~\eqref{C3BDF3} between $F_3(\tau)$ and $C_3(\tau)$, the maximum average work of assistance extractable by any classical LHS ensemble under a three-setting  scheme is strictly upper-bounded by:
\begin{equation}\label{wasteerineq3_revised}
 AW^{B|A}_a(\rho_{AB}) \le \frac{1}{\beta}\left(S(\rho_B||\gamma_B) + 1-h\left(\frac{1+1/\sqrt{3}}{2}\right)\right).
\end{equation}
This  threshold holds because $F_3 \le 1$  for any valid LHS ensemble, and the three-setting linear steering parameter $F_3(\rho_{AB})$ for an arbitrary bipartite state is  lower-bounded by the canonical Bell-diagonal state $\tau'$ obtained via cooperative local filtering operations~\cite{Ducuara2024hidden}. 
An experimental or theoretical extraction of average work---obtained when Alice performs local measurements along symmetric triads of mutually unbiased directions that optimize three-setting simultaneous correlations ($C_3$)---that strictly violates the upper bound of Eq.~(\ref{wasteerineq3_revised}) certifies a one-sided device-independent quantum enhancement, demonstrating that $C_3$ acts as a direct thermodynamic fuel for the triple-basis quantum Maxwell's demon.

\subsubsection{1SSDI steering as a resource}

Let us now examine the three-setting steering scenario with the dimension of Alice's system bounded to be that of a qubit ($d_A = 2$). Within this paradigm, we analyze whether three-setting 1SSDI steering can be exploited to provide a more robust quantum thermodynamic advantage than the two-setting setup. The average work of assistance is upper-bounded by:
\begin{align}\label{waunsteerineq3}
 AW^{B|A}_a(\rho_{AB}) \le \frac{1}{\beta}\left(S(\rho_B||\gamma_B) + 1-h\left(\frac{1+1/3}{2}\right) \right),
\end{align}
for CQ states that do not exhibit 1SSDI steering since $F_3 \le 1/3$ for all separable states~\cite{PMS+16}. Comparing Eqs.~(\ref{wasteerineq3_revised}) and (\ref{waunsteerineq3}), it follows that certain unsteerable Bell-diagonal states which have 1SSDI steering—and cannot be used to provide a quantum advantage in the two-setting scenario—can successfully demonstrate a quantum advantage under this three-setting framework. In this case, $C_3(\tau)$ precisely quantifies the quantum advantage using a Bell-diagonal state as in the case of the inequality (\ref{wasteerineq3_revised}).

When Alice operates in three MUBs where one is a $C_1$-basis, the average work of assistance becomes a function of all three SCMUB parameters ($C_1$, $Q_2$, and $Q_3$). Structurally extending the two-setting derivation, the selection of this triple-complementary configuration is physically motivated by the fact that $Q_3$ uniquely quantifies this three-setting 1SSDI resource capability~\cite{WMC+14,GW14,Jeb+25}. In this case, the average work of assistance is given by:
\begin{equation}
\label{AW_1SSDI_three_setting}
\begin{split}
AW^{B|A}_a(\rho_{AB})  = \frac{1}{\beta} \bigg[ S(\rho_B||\gamma_B) + \frac{1}{1+c+d} \Big( C_1(\rho_{AB}) \\
+ c Q_2(\rho_{AB}) + d Q_3(\rho_{AB}) \Big) \bigg].
\end{split}
\end{equation}
For qubit-Alice CQ states lacking 1SSDI steering, MUB complementarity forces $Q_2 = Q_3 = 0$ while $C_1 \le 1$, limiting the average work via:
\begin{align}\label{WaQ3steer}
    AW^{B|A}_a(\rho_{AB}) \le \frac{1}{\beta}\left(S(\rho_B||\gamma_B) + \frac{1}{1+c+d} \right).
\end{align}
Equation~\eqref{WaQ3steer} represents a 1SSDI steering inequality detecting three-setting dimension-bounded nonclassicality. The rigorous mathematical derivation of this three-setting boundary collapse, dictated by the concurrent optimization over triple-complementary MUB measures, is detailed in~\ref{app:derivation_three_setting}. 

An analogous physical and operational contrast emerges when comparing the three-setting 1SSDI steering inequalities presented in Eq.~\eqref{waunsteerineq3} and Eq.~\eqref{WaQ3steer}. While both formulations successfully witness dimension-bounded nonclassicality under three complementary measurement configurations, they exploit the underlying SCMUB resource structure in fundamentally different ways. Inequality~\eqref{waunsteerineq3} establishes a static, symmetric classical threshold based on the simultaneous correlation $C_3$. Conversely, Eq.~\eqref{WaQ3steer} presumes a directed, adaptive optimization strategy where Alice’s primary observable is aligned with the optimal $C_1$-basis, forcing the remaining channels to trace the residual quantum correlation metrics $Q_2$ and $Q_3$ explicitly. 

Crucially, while the constant classical threshold of Eq.~\eqref{waunsteerineq3} remains fixed at $1 - h(2/3) \approx 0.0817$, the dynamic boundary of Eq.~\eqref{WaQ3steer} scales as $\frac{1}{1+c+d}$. Equating these two boundaries resolves the exact multi-setting threshold constraint equation $c + d = \frac{h(2/3)}{1-h(2/3)} \approx 11.24$. In highly asymmetric multi-setting pools where the joint ensemble distribution weights satisfy $c+d > 11.24$ (which reduces symmetrically to $c > 10.24$ when $d = 1$), this dynamic boundary collapses significantly below the static symmetric baseline of Eq.~\eqref{waunsteerineq3} ($\approx 0.0817$). Consequently, Eq.~\eqref{WaQ3steer} provides a substantially more sensitive thermodynamic witness, effectively broadening the operational domain of unsteerable states that can be certified as a nonclassical fuel for a three-setting quantum Maxwell's demon.

\begin{figure}[!t]
    \centering
    \includegraphics[width=0.72\textwidth]{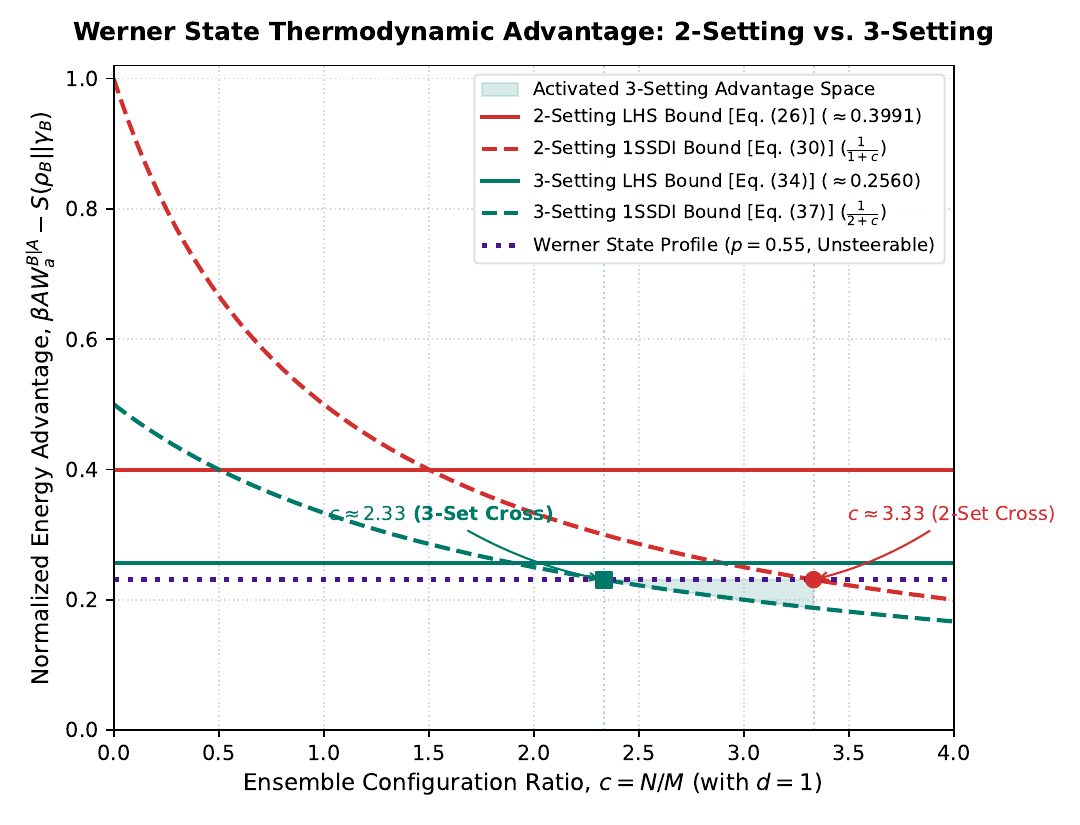}
    \caption{Comparison of the normalized average work enhancement $\beta AW_a^{B|A} - S(\rho_B||\gamma_B)$ for an unsteerable bipartite Werner state ($p = 0.55$) in two-setting and three-setting scenarios. Under two measurement configurations, the state can only certify a dimension-bounded 1SSDI advantage when the ensemble ratio exceeds $c \approx 3.33$, corresponding to the intersection with the decaying threshold $\frac{1}{1+c}$. However, when moving to the three-setting scenario (setting $d = 1$), the dimension-bounded classical benchmark drops to $\frac{1}{1+c+d} = \frac{1}{2+c}$, shifting the critical crossover validation threshold back significantly to $c \approx 2.33$. The shaded teal domain defines the newly activated three-setting advantage space, providing explicit proof that maximizing the number of complementary measurement configurations noticeably broadens the domain of states capable of showcasing nonclassical quantum thermodynamic utility under dimensional trust.}
    \label{fig:werner_advantage}
\end{figure}

To showcase how migrating from a two-setting configuration to a triple-complementary MUB setup structurally broadens this advantage space, a comparative tracking using the isotropic bipartite Werner state family ($\rho_W$) is presented in Fig.~\ref{fig:werner_advantage}. This family of states is defined as a convex combination of the maximally entangled pure Bell state and the completely mixed unpolarized background:
\begin{equation}
\label{eq:werner_state_def}
\rho_W = p \vert \Phi^+ \rangle \langle \Phi^+ \vert + (1-p)\frac{\openone_4}{4},
\end{equation}
where $p \in [0,1]$ denotes the visibility parameter, and $\vert \Phi^+ \rangle = \frac{1}{\sqrt{2}}(\vert 00 \rangle + \vert 11 \rangle)$ is the standard maximally entangled Bell state. For this isotropic family, the correlation vector components are uniform, satisfying $|t_1| = |t_2| = |t_3| = p$. We evaluate a  unsteerable Werner state with visibility $p = 0.55$, meaning it is structurally incapable of violating either the two-setting ($F_2 \le 1$) or three-setting ($F_3 = \sqrt{3}p \approx 0.9526 \le 1$) unconstrained LHS bounds.

When restricted to two measurement configurations ($d=0$), the state's flat normalized advantage profile ($C_1 = Q_2 = 1 - h(0.775) \approx 0.2308$) can only cross the decaying 1SSDI threshold ($\frac{1}{1+c}$) in the highly asymmetric regime where $c > 3.33$. However, upon introducing the third MUB setting under a representative symmetric pool weight ($d = 1$), the general dimension-bounded classical benchmark expression collapses to $\frac{1}{1+c+d} = \frac{1}{2+c}$. This structural shift tightens the semi-device-independent boundary, causing the critical crossover validation threshold requirement to migrate back significantly from $c \approx 3.33$ down to $c \approx 2.33$.

The shaded teal domain highlighted in Fig.~\ref{fig:werner_advantage} explicitly isolates this newly activated three-setting advantage space. As visually traced in the profile curves, the unsteerable Werner state candidate at $p = 0.55$ remains trapped entirely beneath both the two-setting ($\approx 0.3991$) and three-setting ($\approx 0.2560$) unconstrained LHS thresholds, yet seamlessly breaches the dynamically scaling 1SSDI curves. This visual gap provides clear proof that maximizing the number of complementary measurement settings noticeably lowers the required ensemble asymmetry needed to extract nonclassical work, allowing Bob's microscale TO to bypass the noise mask of the mixed unpolarized background. Consequently, a significantly wider class of unsteerable states can be actively engineered to fuel microscopic quantum Maxwell's demons under dimensionally trusted scenarios.

\section{Conclusions and Discussions}

In this work, we have established a rigorous operational bridge between the resource theory of quantum thermodynamics and the foundational characterization of nonclassical correlations. Specifically, we analyzed the framework of assisted work distillation originally introduced by Morris \etal~\cite{MLA19}, expanding its scope to asymmetric setups involving multiple local measurement configurations on the untrusted side. By optimizing Alice’s POVMs within sets of MUBs, we proved that the simultaneous presence of correlations in complementary bases serves as an explicit, quantifiable resource capable of fueling thermodynamic advantages for a localized observer.

\textbf{Core Physical Insights:}
\begin{itemize}
    \item \textbf{Thermodynamic Crossover Mechanism:} The operational boundaries of unconstrained LHS configurations remain static. However, when the preparing party is restricted to a trusted qubit dimension ($d_A = 2$), MUB complementarity across complementary measurement settings forces the dimension-bounded classical thresholds to decay dynamically as a function of the ensemble storage pool capacity weights ($c$ and $d$).
    \item \textbf{Activation of Hidden Utility:} Unconstrained EPR steering is a sufficient but entirely non-necessary condition for quantum work extraction. Weakly correlated entangled states that are unsteerable under standard scenarios—such as the isotropic bipartite Werner state at an intermediate visibility landscape ($p = 0.55$)—can comfortably breach our dimension-bounded thresholds. This proves that unsteerable assemblages still house a hidden thermodynamic utility that can be actively certified simply by asserting dimensional trust.
    \item \textbf{Multi-Basis Scaling Advantages:} Increasing the number of available complementary measurement configurations (transitioning from a two-setting to a three-setting configuration) dramatically lowers the critical pool asymmetry required to unlock a verified nonclassical quantum advantage.
\end{itemize}

\textbf{Practical Architectural Pathways:}
From a practical perspective, these findings open up a highly viable, device-independent pathway for implementing nanoscale work extraction protocols using current, dimensionally restricted quantum architectures. Because our inequalities certify nonclassical thermodynamic utility without requiring a fully trusted or calibrated measurement apparatus on the preparing side, they are highly robust against local detection noise and misalignment. This makes our multi-setting adaptive scheme natively compatible with state-of-the-art superconducting circuits or integrated photonic platforms, where physical qubits naturally provide the required 1SSDI dimensional bounding ($d_A = 2$) to unlock and exploit hidden quantum thermodynamic resources.
This architectural compatibility is highlighted by recent experimental setups utilizing programmable superconducting quantum processors to demonstrate the exact interplay between Maxwell demons and EPR steering frameworks~\cite{WangDemonSteering2024}.

Looking ahead, a natural extension of this multi-setting thermodynamic paradigm lies in generalizing these work-based certification tools to high-dimensional systems ($d_A > 2$) or multipartite configurations. Investigating how genuine multipartite entanglement hierarchies scale under asymmetric storage distributions could provide key insights into the ultimate bounds of macroscopic quantum thermodynamics.

\begin{ack}
The authors acknowledge that the initial framework of this research dates back to 2020, when C.J. presented these core concepts as a poster at the Young Researchers Forum on Quantum Information Science 2020 in Hsinchu. C.J. and R.-K. L. thank Chung-Yun Hsieh for helpful and stimulating discussions, and A.~de~Oliveira~Junior for helpful correspondence and for bringing his related work to our attention. C.J. deeply thanks H.-S. Goan for serving as a postdoctoral supervisor, for his foundational guidance in defining the broader research project framework, and for insightful discussions that helped connect abstract quantum foundations to practical thermodynamic applications. C.J. also acknowledges support as a postdoctoral researcher hosted by the Center for Quantum Science and Engineering (CQSE) at National Taiwan University.
\end{ack}

\subsection*{Data Availability Statement}
No new data were created or analysed in this study. All analytical derivations and numerical evaluations supporting the findings of this article are included within the manuscript.

\subsection*{Conflict of Interest}
The authors declare that there are no conflicts of interest or competing financial interests related to this work.

\subsection*{Author Contributions}
\textbf{Chellasamy Jebarathinam:} Conceptualization, Methodology, Formal analysis, Investigation, Writing -- original draft, Writing -- review \& editing. Approached R.-K.~L. to initiate the collaboration, proposed the core research problem, developed the theoretical framework, performed the analytical derivations and numerical evaluations, and drafted the manuscript. \\
\textbf{Debarshi Das:} Methodology, Formal analysis, Validation, Writing -- review \& editing. As a close collaborator on steering and superunsteerability, contributed to the derivation of the work-based inequalities and their physical connection to quantum Maxwell's demons. \\
\textbf{Ray-Kuang Lee:} Conceptualization, Supervision, Project administration, Validation, Writing -- review \& editing. Invited C.~J. to collaborate following C.~J.'s initial approach, supervised the project, and helped establish the connection to quantum Maxwell's demons through in-person office discussions, a seminar, and a poster presentation.

\appendix

\section{Formal Derivation of the Two-setting 1SSDI Steering Inequality and Boundary Collapse}
\label{app:derivation}

To establish the validity of the dimension-bounded validation boundary presented in Eq.~\eqref{WaQ2steer}, we determine the maximum value that the weighted correlation resource sum can achieve over all configurations devoid of 1SSDI nonclassicality. For an arbitrary bipartite quantum state $\rho_{AB}$, we define this correlation resource sum $\mathcal{R}(\rho_{AB}, c)$ across the two complementary measurement bases as:
\begin{equation}
\label{eq:appA_resource_sum}
\mathcal{R}(\rho_{AB}, c) = \frac{C_1(\rho_{AB}) + c Q_2(\rho_{AB})}{1+c},
\end{equation}
where $c = N/M \ge 0$ represents the non-negative capacity distribution ratio between Bob's independent storage particle pools. Because every LHS model for a qubit-bounded untrusted party ($d_A = 2$) can be reproduced by a  CQ state $\rho_{CQ}$ of the form in Eq.~\eqref{rhoCQ_def} with $d_\lambda \le 2$, the strict classical benchmark is given by maximizing Eq.~\eqref{eq:appA_resource_sum} directly over the set of all qubit-Alice CQ states:
\begin{equation}
\label{eq:appA_optimization_condensed}
\mathcal{R}_{\max}(c) = \max_{\rho_{CQ}} \left[ \frac{C_1(\rho_{CQ}) + c Q_2(\rho_{CQ})}{1+c} \right].
\end{equation}

Any qubit-Alice CQ state $\rho_{CQ}$ can be written in its canonical spectral decomposition on subsystem $A$ as:
\begin{equation}
\label{eq:appA_CQ_decomp}
\rho_{CQ} = p \proj{0}_A \otimes \rho_{B|0} + (1 - p) \proj{1}_A \otimes \rho_{B|1},
\end{equation}
where $p \in [0,1]$, $\{\ket{0}_A, \ket{1}_A\}$ is an orthonormal basis on Alice's qubit, and $\rho_{B|0}, \rho_{B|1}$ are normalized quantum states on Bob's subsystem. We evaluate $C_1(\rho_{CQ})$ and $Q_2(\rho_{CQ})$ directly across two exhaustive cases:

\begin{enumerate}
    \item \emph{Trivial product case ($\rho_{B|0} = \rho_{B|1} = \rho_B$, or $p \in \{0,1\}$):} Here $\rho_{CQ} = \rho_A \otimes \rho_B$ is a product state, so every local measurement on $A$ leaves Bob's conditional state invariant ($\tilde{\rho}_{B|i} = \rho_B$). Consequently, the Holevo information vanishes in every basis, yielding $C_1(\rho_{CQ}) = 0$ and $Q_2(\rho_{CQ}) = 0$, and hence $\mathcal{R}(\rho_{CQ}, c) = 0$.

    \item \emph{Non-product CQ case ($\rho_{B|0} \neq \rho_{B|1}$ and $0 < p < 1$):} By the strict concavity of the von Neumann entropy, the Holevo quantity $\chi\{\rho_{CQ} \mid \{\ket{a_i}_A\}\}$ in Eq.~\eqref{eq:holevo_unified} is uniquely maximized when Alice measures in the eigenbasis $\{\ket{0}_A, \ket{1}_A\}$ of $\rho_{CQ}$ on $A$, since any other projective basis produces conditional states on $B$ that are nontrivial convex mixtures of $\rho_{B|0}$ and $\rho_{B|1}$, strictly reducing the Holevo information. Thus, the $C_1$-basis in Eq.~\eqref{defC1} is uniquely fixed (up to global phases) as $\{\ket{0}_A, \ket{1}_A\}$, and the classical correlation is bounded by the binary Shannon entropy:
    \begin{equation}
    \label{eq:appA_C1_bound}
    C_1(\rho_{CQ}) = \chi\bigl(\{p, \rho_{B|0};\, 1-p, \rho_{B|1}\}\bigr) \le h(p) \le 1,
    \end{equation}
    with equality $C_1(\rho_{CQ}) = 1$ achieved if and only if $p = 1/2$ and $\rho_{B|0} \perp \rho_{B|1}$ (for instance, $\rho_{CQ}^{\text{max-}C_1} = \frac{1}{2}\proj{0}_A \otimes \proj{0}_B + \frac{1}{2}\proj{1}_A \otimes \proj{1}_B$).

    Next, to evaluate $Q_2(\rho_{CQ})$ in Eq.~\eqref{defQ2}, Alice must measure in a second basis $\{\ket{a_j^2}_A\}_{j=0}^{1}$ that is mutually unbiased with respect to the unique $C_1$-basis $\{\ket{0}_A, \ket{1}_A\}$. By definition of MUBs in $d_A = 2$, the overlap probabilities satisfy $|\braket{0|a_j^2}|^2 = |\braket{1|a_j^2}|^2 = 1/2$ for both outcomes $j \in \{0, 1\}$. The outcome probabilities on Alice's side are therefore $q_j = \Tr[(\proj{a_j^2}_A \otimes \openone_B)\rho_{CQ}] = \frac{1}{2}p + \frac{1}{2}(1-p) = \frac{1}{2}$, and the corresponding conditional states on Bob's side collapse identically to Bob's reduced state $\rho_B$:
    \begin{align}
    \label{eq:appA_cond_collapse}
    \tilde{\rho}_{B|j} &= \frac{1}{q_j} \Tr_A \left[ \left( \proj{a_j^2}_A \otimes \openone_B \right) \rho_{CQ} \right] \nonumber \\
    &= 2 \left( \frac{p}{2}\rho_{B|0} + \frac{1-p}{2}\rho_{B|1} \right) = p\rho_{B|0} + (1-p)\rho_{B|1} = \rho_B.
    \end{align}
    Because $\tilde{\rho}_{B|0} = \tilde{\rho}_{B|1} = \rho_B$, Bob's accessible Holevo information along any MUB complementary to the $C_1$-basis vanishes identically:
    \begin{align}
\label{eq:main_text_monogamy_explicit}
\chi\big\{\rho_{CQ}\big|\{\ket{a_j^2}_A\}\big\} &= S(\rho_B) - \frac{1}{2}S(\rho_B) - \frac{1}{2}S(\rho_B) = 0 \nonumber \\
&\implies Q_2(\rho_{CQ}) = 0.
\end{align}
\end{enumerate}
This result formalizes the MUB complementarity constraint for qubit-Alice CQ states: because $\rho_{CQ}$ is diagonal in the $C_1$-basis, it cannot sustain nonvanishing Holevo correlations along any mutually unbiased basis ($Q_2(\rho_{CQ}) = 0$).

Combining both cases, every qubit-Alice CQ state satisfies $C_1(\rho_{CQ}) \le 1$ and $Q_2(\rho_{CQ}) = 0$ simultaneously. Substituting these exact constraints into Eq.~\eqref{eq:appA_optimization_condensed} isolates the maximum resource capacity achievable by any valid CQ state ensemble across all non-negative storage pool distributions ($c \ge 0$):
\begin{equation}
\label{eq:appA_Rmax}
\mathcal{R}_{\max}(c) = \max_{\rho_{CQ}} \left[ \frac{C_1(\rho_{CQ}) + c(0)}{1+c} \right] = \frac{1}{1+c}.
\end{equation}
Substituting this optimized resource upper bound back into the master average work mapping in Eq.~\eqref{AW_1SSDI_two_setting} directly yields the dynamically scaling 1SSDI steering inequality presented in Eq.~\eqref{WaQ2steer}. This completes the proof of the dynamic boundary collapse.

\section{Formal Derivation of the Three-Setting 1SSDI Steering Inequality and Boundary Collapse}
\label{app:derivation_three_setting}

In this Appendix, we extend the CQ-state proof of~\ref{app:derivation} to establish the three-setting 1SSDI steering inequality presented in Eq.~\eqref{WaQ3steer}, parameterized by the non-negative storage pool capacity weights $c = N/M \ge 0$ and $d = O/M \ge 0$.

For an arbitrary bipartite quantum state $\rho_{AB}$, the generalized multi-setting correlation resource sum distributed across three distinct MUBs is defined as:
\begin{equation}
\label{eq:appB_resource_sum}
\mathcal{R}(\rho_{AB}, c, d) = \frac{C_1(\rho_{AB}) + c Q_2(\rho_{AB}) + d Q_3(\rho_{AB})}{1+c+d}.
\end{equation}
To isolate the strict classical threshold that bounds dimensionally restricted configurations, we maximize this objective function over the set of all qubit-Alice CQ states $\rho_{CQ}$ given in Eq.~\eqref{eq:appA_CQ_decomp}:
\begin{equation}
\label{eq:appB_optimization_condensed}
\mathcal{R}_{\max}(c, d) = \max_{\rho_{CQ}} \left[ \frac{C_1(\rho_{CQ}) + c Q_2(\rho_{CQ}) + d Q_3(\rho_{CQ})}{1+c+d} \right],
\end{equation}
where the optimization domain is restricted to the set of CQ states because every LHS model can be reproduced by a CQ state representation.

From the definitions of $Q_2$ in Eq.~\eqref{defQ2} and $Q_3$ in Eq.~\eqref{defQ3}, the third measurement basis $\{\ket{a_k^3}_A\}_{k=0}^1$ is constrained to be mutually unbiased with respect to both the $C_1$-basis and the $Q_2$-basis, whereas $Q_2$ optimizes over all bases mutually unbiased to the $C_1$-basis alone. Because the Holevo quantity is non-negative and $Q_3$ maximizes over a strictly smaller subset of admissible MUBs, the ordering $0 \le Q_3(\rho_{AB}) \le Q_2(\rho_{AB})$ holds universally. Since~\ref{app:derivation} proves that $C_1(\rho_{CQ}) \le 1$ and $Q_2(\rho_{CQ}) = 0$ for every qubit-Alice CQ state, it follows immediately that:
\begin{equation}
\label{eq:appB_monogamy_condensed}
Q_2(\rho_{CQ}) = 0 \implies Q_3(\rho_{CQ}) = 0.
\end{equation}

Substituting $C_1(\rho_{CQ}) \le 1$ and $Q_2(\rho_{CQ}) = Q_3(\rho_{CQ}) = 0$ into Eq.~\eqref{eq:appB_optimization_condensed} yields the maximum classical resource capacity for all non-negative joint distribution weights ($c \ge 0, d \ge 0$):
\begin{equation}
\label{eq:appB_Rmax}
\mathcal{R}_{\max}(c, d) = \max_{\rho_{CQ}} \left[ \frac{C_1(\rho_{CQ}) + c(0) + d(0)}{1+c+d} \right] = \frac{1}{1+c+d}.
\end{equation}
Substituting this optimized resource upper bound back into the master three-setting work relation given in Eq.~\eqref{AW_1SSDI_three_setting} directly resolves the certified 1SSDI steering inequality in Eq.~\eqref{WaQ3steer}:
\begin{equation}
\label{eq:appB_final}
AW_a^{B|A} \le \frac{1}{\beta}\left(S(\rho_B||\gamma_B) + \frac{1}{1+c+d} \right).
\end{equation}
This completes the proof of the multi-setting dynamic boundary collapse.

\section*{References}
\bibliographystyle{iopart-num}
\bibliography{QT}

\end{document}